\documentclass[english,letter,12pt,twosided]{article}
\usepackage{UF_FRED_paper_style}

\usepackage{lipsum}  
\usepackage{enumitem}
\usepackage{tikz}
\usepackage{mathrsfs}
\usepackage{threeparttable}
\usepackage{amsmath}
\usepackage{booktabs}
\usepackage{bookmark}
\usepackage{caption}
\usepackage{graphicx}
\usepackage{enumitem}
\usepackage[T1]{fontenc}
\usepackage[utf8]{inputenc}
\usepackage{longtable}

\usepackage{amsthm}

\theoremstyle{plain}       
\newtheorem{theorem}{Theorem}

\theoremstyle{definition}  

\theoremstyle{remark}      

\usepackage{datatool}

\DTLloaddb[keys={title,value}]{stats}{paper_stats.csv}

\definecolor{PennBlue}{RGB}{001,031,091}
\definecolor{PennRed}{RGB}{153,0,0}
\hypersetup{
pdfborder = {0 0 0},
    colorlinks,
    citecolor=PennRed,
    filecolor=PennRed,
    linkcolor=PennRed,
    urlcolor=PennRed}

\title{\textbf{Sector-Specific Substitution and the Effect of Sectoral Shocks}
\thanks{I am grateful to my advisors Lawrence Christiano, Benjamin Jones, Kunal Sangani, and Alireza Tahbaz-Salehi, for their guidance. I also thank Aaron Amburgey, Vivek Bhattacharya, Martin Eichenbaum, Joseph Ferrie, Gaston Illanes, Diego Känzig, José Luis Lara, Giorgio Primiceri, Ramya Raghanavan, Matthew Rognlie, and Dalton Zhang, for their helpful comments.}}

\author{Jacob Toner Gosselin\\
    \href{mailto:jacob.gosselin@u.northwestern.edu}{Northwestern University} 
    }
    
\date{\today}

\begin{document}
\begin{filecontents*}{paper_stats.csv}
"title","value"
"low-sd-year",0.265
"high-sd-year",1.127
"median-sd-year",0.631
"median-yoy-change",0.686
"low-sd-Code",0.442
"high-sd-Code",1.654
"median-sd-Code",0.751
"corr-stat",0.417
\end{filecontents*}

\captionsetup{justification=centering, singlelinecheck=false}
\vfill
{\setstretch{.8}
\maketitle
\begin{abstract}

How a shock to an individual sector propagates to the prices of other sectors and aggregates to GDP depends on how easily sectoral goods can be substituted in production, which is determined by the intermediate input substitution elasticity. Past estimates of this parameter in the US have been restrictive: they have assumed a common elasticity across industries, and have ignored the use of imports in production. This paper uses a novel empirical strategy to produce new estimates without these restrictions, by exploiting variation in import ratios and input expenditure shares from the BEA Input-Output Accounts.  I find that sectors differ meaningfully in their ability to substitute inputs in production, and that the uniform estimate of the intermediate input substitution elasticity is biased downwards relative to the median sector-specific estimate. Relative to imposing the uniform elasticity, sector-specific substitution causes domestic prices to rise more in response to oil import shocks and less in response to semiconductor import shocks. It also implies the average GDP response to a sectoral business cycle is 0.35\% higher, making sectoral business cycles 17.7\% less costly.

\end{abstract}
}
\vfill


\section*{Introduction}

Over the past fifteen years, a large literature has emerged documenting the importance of input-output linkages between sectors for the micro-to-macro propagation of shocks \citep{acemogluNetworkOriginsAggregate2012, gabaixGranularOriginsAggregate2011}. In this literature, production function parameters governing the substitutability of inputs play a central role. Shocks to individual sectors can have a large effect on aggregate output if the shocked sectors are important suppliers to industries who are unable to substitute inputs (\cite{baqaeeMacroeconomicImpactMicroeconomic2019}). 

Consequentially, these parameters are central in many policy debates. Consider the discussion that unfolded in Germany following the Russian invasion of Ukraine, where economists put forth competing estimates of the macro effect of cutting off Russian energy imports. Their estimates rested almost entirely on the elasticity of substitution between Russian oil and other inputs in the production of German goods. Small differences in its calibration ballooned the projected impact on German GDP (\cite{bachmannWhatIfMacroeconomic2024}).

Despite their importance, the estimates of these parameters used to calibrate multi-sector models have been restrictive. Existing estimates for the United States are obtained under the assumptions that (a) the elasticity of substitution is common across sectors, and (b) the economy is closed.\footnote{Recent work by \citet{poirierReallocationDynamicsProduction2025} allows elasticities to vary across 32 major US industries. I discuss how my work differs in detail below.} 

In this paper I relax those assumptions, estimating sector-specific intermediate input substitution elasticities that account for import use in production, for 66 US industries. I find that industries differ meaningfully in their ability to substitute inputs in production, and that the uniform estimate of the intermediate input substitution elasticity is biased downwards relative to the median sector-specific estimate. I quantify these results in a multi-sector general equilibrium (GE) model calibrated to the US economy. Relative to imposing the uniform elasticity, sector-specific substitution causes domestic prices to rise more in response to oil import shocks and less in response to semiconductor import shocks, and lowers the GDP cost of sectoral business cycles.  

I consider a standard multi-sector GE model a la \cite{horvathSectoralShocksAggregate2000}, where sector-specific goods are produced with a nested-CES production technology, to be consumed by the household and used in the production of other goods. I introduce imported varieties of tradeable goods as inputs used in production under the Armington assumption. I derive the first-order response of prices and the second-order response of GDP to exogenous shocks; both equations depend on the intermediate input substitution elasticity, my parameter of interest. I emphasize how these expressions change in the presence of sectoral heterogeneity in this parameter.

I estimate sector-specific intermediate input substitution elasticities using changes in expenditure shares constructed from the Bureau of Economic Analysis (BEA) Input-Output accounts. These changes provide variation at the industry-input-year level, for 66 major US industries. Past work has not fully leveraged this feature of the data, limiting identifying variation due to endogeneity concerns \citep{atalayHowImportantAre2017}. Because I solely focus on parameters related to intermediate inputs, I am able to isolate variation in expenditure share shifts \textit{within} industries and across inputs to address endogeneity concerns without relying on an instrument. This variation allows me to identify the intermediate input substitution elasticity with more precision than previous work.

To account for import use in production, I derive an empirical specification relating changes in expenditure shares to changes in domestic sectoral prices and changes in import ratios. This reduced form is non-linear in the intermediate input substitution elasticity and the Armington elasticity between domestic and imported varieties of sectoral goods. I estimate both elasticities jointly using generalized method of moments (GMM). My estimates of the intermediate input substitution elasticity exhibit meaningful sectoral heterogeneity. They also reveal that the uniform estimate previously used in the literature is biased downwards relative to the median sector-specific estimate, due to the high variation of expenditure shares in sectors related to oil and gas production.

To assess the quantitative relevance of these results, I calibrate my multi-sector GE model to the US economy and simulate the response of prices and GDP to import price shocks and sectoral productivity shocks. I find that sector specific substitution raises the price response of ``Support activities for mining'' to an oil import by an order of magnitude of 27.4\%, and lowers the price response of ``Motor vehicles'' to a semiconductor import shock by an order of magnitude of 30.2\%. Sector-specific substitution also implies the average GDP response to a sectoral business cycle is 0.35\% higher, making sectoral business cycles 17.7\% less costly. 

\paragraph{Related Literature}

My paper contributes to the modern production network literature, which emphasizes how input-output linkages between producers can propagate idiosyncratic shocks \citep{carvalhoProductionNetworksPrimer2019,baqaeeMicroPropagationMacro2023}. One of the key findings in this literature is that the elasticity of substitution plays a key role in determining the nature and extent of such propagation. I provide industry-level estimates of the elasticity of substitution across intermediate inputs, finding that it varies meaningfully across sectors, and that this heterogeneity alters the effect of sectoral shocks. My estimates are thus relevant to work quantifying the macro effect of sectoral shocks in the US using multi-sector models of production \citep{atalayHowImportantAre2017, baqaeeMacroeconomicImpactMicroeconomic2019,carvalhoSupplyChainDisruptions2021}.

My empirical strategy is closely related to the work of \citet{atalayHowImportantAre2017}. Atalay also uses industry-input expenditure shares constructed from the BEA to identify his parameters of interest. However, because he aims to identify all parameters of sectoral production, he is forced to rely on instrumental variables, which reduce his identifying variation.\footnote{\citet{miranda-pintoFlexibilityFrictionsMultisector2022} use the same instrument regression.}  Because I solely focus on the intermediate input substitution elasticity, I do not need to rely on instrumental variables to avoid endogeneity. Instead, I use variation \textit{within} industry-years and across inputs to identify my parameters.\footnote{A more in-depth discussion of this point can be found in Section \ref{subsec:reduced_form} and Appendix \ref{ap:atalay}.} This greatly expands my identifying variation, which allows me to estimate the intermediate input substitution elasticity at the industry-level. I also use changes in the import ratio to account for import use in production, rather than assume a closed economy. For several industries, this leads to meaningfully different estimates. 

My work is also related to the independent work of \citet{poirierReallocationDynamicsProduction2025}, who estimate sector-varying elasticities across intermediate inputs for 32 US industries between 1997 and 2018. Though we aim to estimate the same parameters, our empirical approaches differ. They use full-information Bayesian methods, which require their full set of equilibrium conditions to be correctly specified. I estimate my parameters via GMM, using an empirical specification based solely on the cost-minimization problem of the representative firm in each sector. My approach produces more granular estimates (66 vs 32 US industries); it also accounts for import use in production, while \citet{poirierReallocationDynamicsProduction2025} assume a closed economy. As discussed in Section \ref{subsec:empirical_results}, our uniform elasticity results line up, but there are some notable differences sector-by-sector, resulting in different predictions for the macro effect of shocks to several industries.

The paper is divided into three sections. Section \ref{sec:theory} presents my theoretical framework: I describe my model, examine the theoretical relevance of my parameters of interest, and derive the empirical specification I use to estimate my parameters. In Section \ref{sec:empirical} I review my data sources and estimation method, and present my empirical results. In Section \ref{sec:discussion} I quantify the economic implications of my empirical estimates in a multi-sector GE model calibrated to the US economy.

\section{Theoretical Framework}
\label{sec:theory}

In this section I consider a standard multi-sector GE model, where sector-specific goods are produced with a nested-CES production technology, to be consumed by the household and used in the production of other goods \citep{horvathSectoralShocksAggregate2000}. I introduce imported varieties of tradeable goods as inputs used in production under the Armington assumption. I treat foreign prices are exogenous, in contrast to richer global trade models \citep{caliendoEstimatesTradeWelfare2015,baqaeeNetworksBarriersTrade2024}. I describe the exogenous parameter of interest in my paper, the substitution elasticity across intermediate inputs $\theta_i$, and describe how sectoral heterogeneity in the parameter mediates the response of sales shares, prices, and GDP to sectoral shocks. I also derive the relation that serves as the basis of my empirical specification, and discuss how it relates to previous work. 

\subsection{Model}
\label{subsec:model}

Consider an economy with $N$ competitive industries producing distinct goods that can be consumed by the household or used in the production of other goods. In every period $t$, each sector $i$ produces output $Y_{it}$ with a nested-CES production technology. The outer nest is a CES aggregator of labor $L_{it}$ and an intermediate input bundle $M_{it}$. The inner nest is a CES aggregator of sector-specific composite goods used as intermediate inputs, $\bar X_{ijt}$. Composite good $\bar X_{ijt}$ is a CES aggregator of domestically-produced good $X_{ijt}$ and imported good $\tilde X_{ijt}$.

\begin{align*}
& Y_{it} = Z_{it} \left(\gamma_i^{\frac{1}{\sigma}} L_{it}^{\frac{\sigma-1}{\sigma}} + (1-\gamma_i)^{\frac{1}{\sigma}} M_{it}^{\frac{\sigma - 1}{\sigma}} \right)^{\frac{\sigma}{\sigma-1}} \\
& M_{it} = \left(\sum_j^N \omega_{ij}^{\frac{1}{\theta_i}} \bar X_{ijt}^{\frac{\theta_i-1}{\theta_i}} \right)^{\frac{\theta_i}{\theta_i-1}} \quad \bar X_{ijt} = \left(\phi_{ij}^{\frac{1}{\xi}} X_{ijt}^{\frac{\xi - 1}{\xi}} + (1-\phi_{ij})^{\frac{1}{\xi}} \tilde X_{ijt}^{\frac{\xi - 1}{\xi}} \right)^{\frac{\xi}{\xi - 1}}
\end{align*}

I assume the elasticity of substitution between the intermediate input bundle and labor ($\sigma$) and the Armington elasticity between imported and domestic variants ($\xi$) are common across sectors, but I allow $\theta_i$, the intermediate input substitution elasticity, to vary across industries. A good $j$ is tradeable if it is imported for at least one industry $i$, i.e. $\exists i \text{ such that } \phi_{ij} < 1$.

The household consumes by maximizing a CES aggregator across domestic sectoral goods, subject to its budget constraint.  

\[
C_t = \max_{\{C_{it}\}} \left(\sum_i^N \beta_i^{\frac{1}{\nu}} C_{it}^{\frac{\nu-1}{\nu}}\right)^{\frac{\nu}{\nu-1}} \text{ subject to } \sum_i^N P_{it} C_{it} = \sum_i W_{it} L_i
\]

I assume labor is a sector-specific factor of production exogenously fixed at an initial value. This is in line with existing empirical literature documenting little to no short-run labor reallocation across sectors in response to sectoral shocks \citep{acemogluImportCompetitionGreat2016}. The labor market is competitive, so wages equal the marginal product of labor in each sector. 

Foreign-currency prices $\{\tilde P_{it}\}_{i=1}^N$ are exogenous. The domestic-currency price of an imported good is $E_t \tilde P_{it}$, where $E_t$ is the real exchange rate (the price of foreign goods in units of the domestic numeraire). Exports are determined by a downward-sloping foreign demand schedule. For tradeable goods foreign demand for the domestic variety is isoelastic in its price (in foreign units):
\[
C^f_{it} = \phi^f_i \left(\frac{P_{it}}{E_t}\right)^{-\tilde \xi}
\]
Where $\phi^f_i$ is a demand shifter and $\tilde \xi$ is the export-demand elasticity. 

The competitive equilibrium is defined in the usual way, as prices and quantities such that all firms make profit-maximizing input choices, the representative household makes utility-maximizing consumption choices, and markets for goods and labor clear.

Expenditure shares of various forms are referenced repeatedly throughout the paper. I define them in Table \ref{tab:expenditure_shares} for ease of reference. 

\begin{table}[h]
\centering
\caption{Expenditure Share Definitions}
\label{tab:expenditure_shares}
\begin{tabular}{ll}
\toprule
\textbf{Definition} & \textbf{Notation} \\
\midrule
Labor Share & $\Gamma_{it} = \frac{W_{it} L_{it}}{P_{it} Y_{it}}$ \\[1em]
Intermediate Expenditure Share & $\Omega_{ijt} = \frac{P_{jt} X_{ijt}}{\sum_k P_{kt} X_{ikt} + E_t \tilde P_{kt} \tilde X_{ikt}}$ \\[1em]
Import Ratio & $\Phi_{ijt} = \frac{P_{jt} X_{ijt}}{P_{jt} X_{ijt} + E_t \tilde P_{jt} \tilde X_{ijt}}$ \\[1em]
Input Output Matrix & $\mathbf A_t := a_{ijt} = \frac{P_{jt} X_{ijt}}{P_{it} Y_{it}}; \; a_{0jt} = \frac{P_{jt} C_{jt}}{\sum_0^N P_{it} C_{it}}, a_{i0t} = 0$ \\[1em]
Leontief Inverse & $\mathbf \Psi_t := \psi_{ijt} = (\mathbb I - \mathbf A_t)^{-1} = \sum^\infty_{k=0} \mathbf A_t^k$ \\
\bottomrule
\end{tabular}
\end{table}

\subsection{The Role of Substitution Elasticities}
\label{subsec:stylized_example}

Recall that substitution elasticities govern the curvature of the CES aggregator isoquants, relating changes in input ratios to changes in the marginal rate of technical substitution (MRTS). As they rise, the isoquants become straight lines, and the CES bundle can be maintained with a wide range of input ratios. As they fall, the isoquants become more curved, and the CES bundle can be maintained only with specific input ratios. In this sense, they govern how easily inputs can be substituted. 

As a consequence, in my model substitution elasticities mediate how sales shares $\lambda_{it} = \frac{P_{it} Y_{it}}{\text{GDP}_t}$ respond to price changes. The proof can be found in Appendix \ref{ap:theory}.

\begin{theorem}
\label{thm:sales_share}
Changes in the sales share of good $i$ is 
\[
\partial \lambda_{it} = \sum_j \sum_k \lambda_{kt} \psi_{jit} a_{kjt} \partial \log a_{kjt} + \sum_j \psi_{jit} NX_{jt}\partial \log NX_{jt}
\]
Where export expenditures are $NX_{jt} = P_{jt} C^f_{jt}$. Changes in the input-output matrix are
\begin{align*}
\partial \log a_{0jt} = a_{0jt}(1-\nu) (\partial \log P_{jt} - \partial \log P_{0t}) \\
\partial \log a_{ijt} = (\sigma - 1) P_{it} + (\theta_i - \sigma) \partial \log Q_{it} + (\xi-\theta_i) \log \bar P_{jt} + (1-\xi) \log P_{jt}
\end{align*}
Changes in CES price indices can be written in terms of changes in sectoral prices as follows:
\[
\partial \log Q_{it} = \sum_j \Omega_{ijt} \partial \log \bar P_{jt}, \quad \partial \log \bar P_{jt} = \Phi_{ijt}  \partial \log P_{jt} + (1 - \Phi_{ijt}) (\partial \log \tilde P_{jt} + \partial \log E_t)
\]
\end{theorem}

By market clearing, changes in sales shares are determined by changes in export revenue and changes in the input-output matrix.\footnote{Changes in export expenditures are $\partial \log NX_{jt} = (1-\tilde \xi) \partial \log P_{jt} + \partial \log E_t$, by construction.} Changes in the input-output matrix are driven by substitution in response to price changes, which is governed by the substitution elasticities. Note that the extent to which substitution by sector $k$ matters for the sales share of sector $i$ depends on its size relative to the economy ($\lambda_{kt}$), and the direct and indirect linkages between its substitution and sector $i$ ($\psi_{jit}$). This \emph{weighting} implies that heterogeneity across sectors in the intermediate input substitution elasticity $\theta_i$ alters how sales shares respond to shocks to first-order.

Through sales shares, substitution elasticities govern how shocks to foreign prices and productivities (1) \emph{propagate} to prices and (2) \emph{aggregate} to GDP. Propagation can be split into two parts: changes in sectoral prices propagate \emph{downstream} to marginal costs through supplier-customer linkages, and changes in industry-to-industry demand propagate \emph{upstream} to sales-shares through customer-supplier linkages. Theorem \ref{thm:sales_share} characterizes upstream propagation, while downstream propagation is characterized below. The proof can be found in Appendix \ref{ap:theory}.

\begin{theorem}
\label{thm:prices}
Changes in prices of sector $i$ are
\[
\partial \log P_{it} = \sum_j \psi_{ijt} (\Gamma_{jt} \partial \log W_{jt} + \sum_k \tilde a_{jkt} \partial \log E_t \tilde P_{kt} - \partial \log Z_{jt})
\]
Changes in wages of sector $i$ are
\[
\partial \log W_{it} = \left(1-\frac 1 \sigma\right) (\partial \log P_{it} + \partial \log Z_{it}) + \frac 1 \sigma \left(\partial \log \lambda_{it}\right)
\]
\end{theorem}

Prices respond to changes in productivities, changes in foreign prices, and changes in sectoral wages through all direct and indirect input-output linkages in production ($\psi_{ijt}$). Sectoral wages respond to changes in sectoral prices, productivities, and sales shares according to the competitive labor market condition. Together, Theorems \ref{thm:sales_share} and \ref{thm:prices} characterize how prices respond to exogenous shocks to foreign prices and productivities. They also imply that \emph{heterogeneity in intermediate input substitution elasticities $\theta_i$ alters the response of prices to import price and productivity shocks to first order.}

The aggregate effect of sectoral productivity shocks is characterized below. The proof can be found in Appendix \ref{ap:theory}.

\begin{theorem}
\label{thm:gdp}
The response of GDP to a sectoral productivity shock to industry $i$ can be approximated to second order as
\[
\partial \log C_t = \lambda_{it} \partial \log Z_{it} + \frac 1 2 \frac{\partial \lambda_{it}}{\partial \log Z_{it}} (\partial \log Z_{it})^2
\]
\end{theorem}

This second-order approximation is a simple extension of Hulten's theorem, which states that GDP is log-linear in sectoral productivity shocks with coefficients corresponding to sales shares. The response of sales shares to sectoral productivity shocks characterizes the non-linear response of GDP to such shocks. Together, Theorems \ref{thm:sales_share}, \ref{thm:prices}, and \ref{thm:gdp} imply that \emph{heterogeneity in intermediate input substitution elasticities $\theta_i$ alters the response of GDP to sectoral productivity shocks to second order.} 

Note that the second-order term is asymmetric, and the direction of the asymmetry depends on substitution elasticities as described in Theorem \ref{thm:sales_share}. When the customers of sector $i$ have sufficiently low substitution elasticities, their expenditure shares move in the same direction as sector $i$'s price. Since sales shares move with expenditure shares, this amplifies[dampens] the GDP effect of negative[positive] shocks to the industry. It also implies an analagous asymmetry for the propagation of price shocks, as described in Theorem \ref{thm:prices}: when prices in sector $i$ rise[fall], the direct and indirect linkages between sector $i$ and $j$ rise[fall], which amplifies[dampens] price propagation to sector $j$. 

In sum, ignoring sectoral heterogeneity in intermediate input substitution elasticities can lead to significant errors in the predicted effect of exogenous shocks on prices and GDP. With that in mind, I now turn to the regression specification I use to estimate these parameters, permitting such heterogeneity.

\subsection{Deriving Empirical Specification}
\label{subsec:reduced_form}

Cost minimization by the representative firm in sector $i$ implies that their expenditure share on sector $j$ is a function of of relative prices, share parameters, and substitution elasticities:

\[
a_{ijt} = \left(\frac{P_{jt}}{\bar P_{jt}}\right)^{1-\xi} \left(\frac{\bar P_{jt}}{Q_{it}}\right)^{1-\theta_i} \phi_{ij} \omega_{ij}
\]

Cost minimization also implies that the CES price index $\bar P_{jt}$ can be written as a function of domestic prices, share parameters, and the import ratio:

\[
\bar P_{jt} = P_{jt} \Phi_{ijt}^{\frac{1}{\xi- 1}} \phi_{ij}^{\frac{1}{1 - \xi}}
\]

Combining these two conditions, taking logs, and taking the total derivative, yields the following result:

\begin{theorem}
\label{thm:reduced_form}
Changes in industry-to-industry expenditure shares are 
\[
\partial \log \Omega_{ijt} = (1-\theta_i) \partial \log P_{jt} + \frac{\xi- \theta_i}{\xi- 1} \partial \log \Phi_{ijt} + \eta_{it}
\]
Where 
\[
\eta_{it} = (\theta_i - 1) \partial \log Q_{it} 
\]
\end{theorem}

Observe that log changes in industry-to-industry expenditure shares $\Omega_{ijt}$ within a given industry-year are log-linear in prices $P_{jt}$ and import ratios $\Phi_{ijt}$, with coefficients corresponding to the intermediate input substitution elasticity $\theta_i$ and the Armington substitution elasticity $\xi$. Theorem $\ref{thm:reduced_form}$ serves as the basis for my empirical specification.

\begin{equation}
\label{eq:reg_spec}
\Delta \log \Omega_{ijt} = (1-\theta_i) \Delta \log P_{jt} + \frac{\xi- \theta_i}{\xi- 1} \Delta \log \Phi_{ijt} + \eta_{it} + \epsilon_{ijt}
\end{equation}

Where $\Delta \log \Omega_{ijt}, \Delta \log P_{jt}, \Delta \log \Phi_{ijt}$ are year-over-year changes in industry-to-industry expenditure shares, prices, and import ratios, while $\eta_{it}$ are industry-year fixed effects. 

Note that in order to control for industry-year fixed effects, industry-input-year variation is required: I cannot rely on data that bundles intermediate inputs together. This eliminates high-quality firm-level microdata for the US (e.g. the US Census Annual Survey of Manufacturers reports bundled intermediate-input expenditures as ``MATCOST''). It is for this reason that, despite their highly aggregated nature, the sector-level data provided by the BEA Input-Output Accounts is uniquely well-equipped to identify my parameters of interest. 

\paragraph{Comparison to Past Work}

As discussed above, \citet{atalayHowImportantAre2017} and \citet{miranda-pintoFlexibilityFrictionsMultisector2022} also estimate substitution elasticities using year-over-year changes in expenditure shares and prices. However, they (1) jointly estimate the intermediate input substitution elasticity $\theta$ and the labor/input bundle substitution elasticity $\sigma$, which reduces their identifying variation, and (2) assume the economy is closed.

On (1), assuming a closed economy, cost minimization implies changes in the input-output matrix are log-linear in changes in prices, with coefficients corresponding to the intermediate input substitution elasticity $\theta$ and the labor/input bundle substitution elasticity $\sigma$:

\[
\partial \log a_{ijt} = (1 - \sigma) \partial \log \left(\frac{Q_{it}}{P_{it}}\right) + (1 - \theta_i) \partial \log \left(\frac{P_{jt}}{Q_{it}}\right) + (\sigma - 1) \partial \log Z_{it} 
\]

\begin{figure}[!h]
    \centering
    \includegraphics[width=\textwidth]{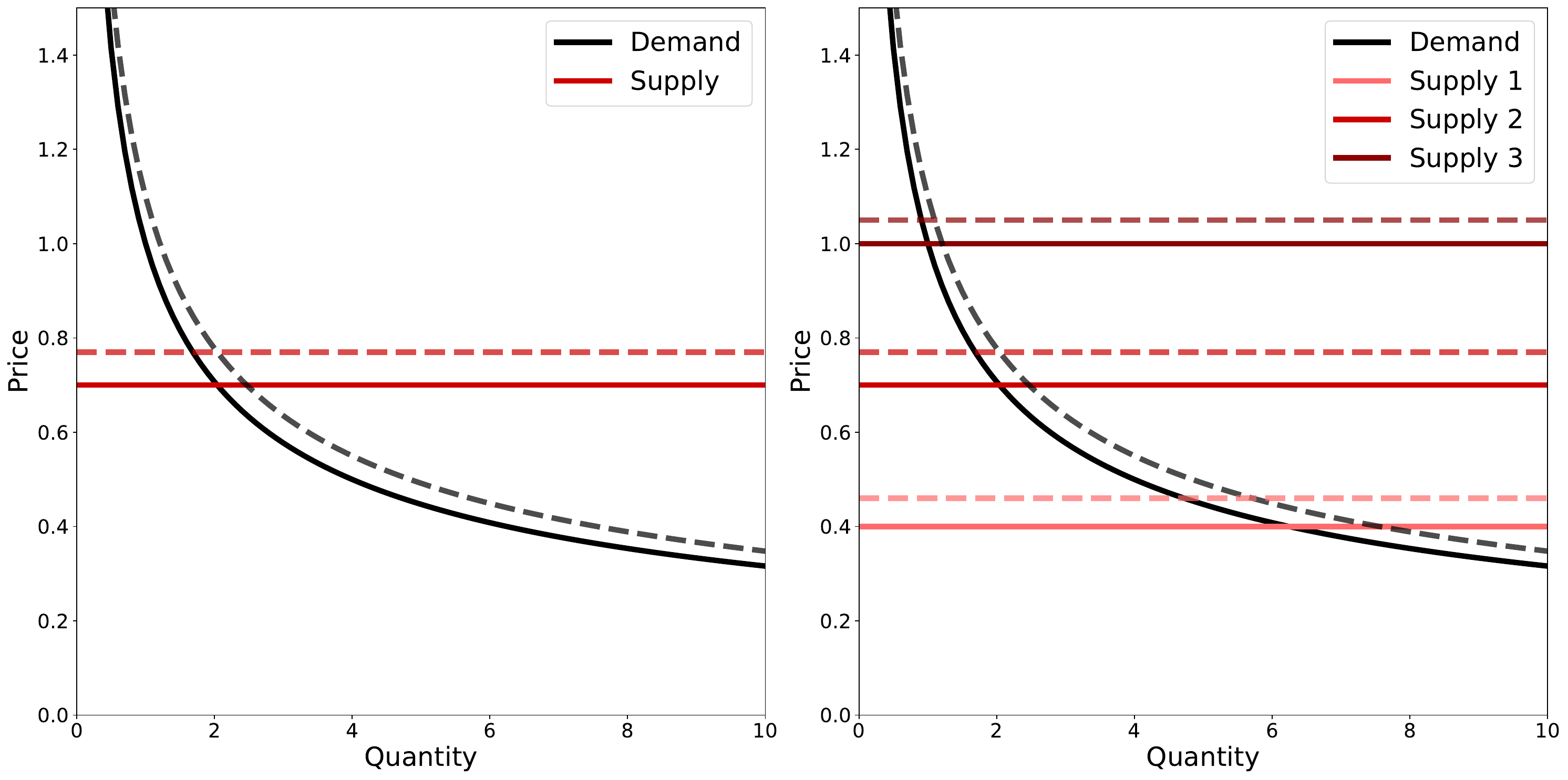}
    \caption{Disentangling Supply and Demand Shifts}
    \label{fig:supply_demand}
\end{figure}

The empirical specification corresponding to this equation suffers from omitted variable bias: changes in productivity $\partial \log Z_{it}$ are unobservable, but correlated with changes in the input-output matrix and changes in relative price $\partial \log \left(\frac{Q_{it}}{P_{it}}\right)$. They also cannot be directly controlled for with industry-year fixed effects, since the variation identifying $\sigma$ is at the industry-year level. This can be understood as an instance of the classic identification problem of disentangling supply and demand shifts, as shown in the left panel of Figure \ref{fig:supply_demand}. Productivity shocks to industry $i$ shift the demand curve, at the same time shocks to relative price $\partial \log \left(\frac{Q_{it}}{P_{it}}\right)$ shift the supply curve. Since both curves move, the demand elasticity cannot be identified. 

\citet{atalayHowImportantAre2017} and \citet{miranda-pintoFlexibilityFrictionsMultisector2022} rely on an instrumental variable approach to overcome this hurdle, reducing their identifying variation and increasing their estimates' standard errors. However, by focusing solely on the substitution elasticity across intermediate inputs $\theta$, I am able to directly control for endogeneity driven by industry-year shocks like $\partial \log Z_{it}$ using industry-year fixed effects. This is because the variation identifying the substitution elasticity across intermediate inputs is at the industry-input-year level, since it governs substitution across \emph{all $j$ intermediate inputs}, rather than just substitution between the labor and the input bundle. Returning to the supply-demand framework, CES implies a common demand elasticity for all goods in the bundle \citep{matsuyamaNonCESAggregatorsGuided2023}. This means we have $j$ supply curves per demand curve, as shown in the right panel of Figure \ref{fig:supply_demand}. Controlling for industry-year fixed effects absorbs the shift in the demand curve, and allows it to be traced by supply curve shifts $\partial \log P_{jt}$.

On (2), if imports are present in production, identifying the substitution elasticity across intermediate inputs using only domestic expenditure shares and domestic prices can bias estimates. Intuitively, this is because changes in industry-to-industry expenditures are driven by two types of substitution: substitution across intermediate inputs, and substitution between domestic and imported varieties. Ignoring the latter can bias estimates of the former. By controlling for changes in the import ratio $\Delta \log \Phi_{ijt}$, I am able to isolate substitution across intermediate inputs, and obtain unbiased estimates of the elasticity governing it. 

For a more in-depth discussion of these points, and a direct comparison of my estimates to those generated using the instrumental variable approach of \citet{atalayHowImportantAre2017}, see Appendix \ref{ap:atalay}. 

\section{Empirical Results}
\label{sec:empirical}

In this section I describe my data construction and estimation strategy, and present my empirical results.

\subsection{Data Construction and Estimation}

To estimate Equation (\ref{eq:reg_spec}), I need measurements of changes in industry-input expenditure shares, import ratios, and prices. I construct these measurements using data from the BEA Input-Output Accounts and GDP by Industry series.\footnote{All BEA data is pulled from their public API.} 

From the Input-Output Accounts I use the ``Domestic Supply of Commodities by Industry - Summary'' and ``Use of Commodities by Industry - Summary'' tables. These tables provide the supply and use of commodities, in \$, by industries at the BEA summary-level (roughly equivalent to 3-digit NAICS classification), from 1997-2024. The Supply table includes the imported value of commodities, in \$, each year.  

I subset the data to only include the 66 non-government summary-level industries. To construct expenditure shares, I sum across commodities the use of commodity $c$ by industry $i$ ($\text{U}_{ict}$) multiplied by the share of commodity $c$ produced by industry $j$ ($\frac{\text{S}_{cjt}}{\sum_k \text{S}_{ckt}}$). To construct import ratios, I sum across commodities the imported share of commodity $c$ ($\frac{\text{Imp}_{ct}}{\sum_k \text{S}_{ckt}}$) times the share of commodity $c$ produced by industry $j$ ($\frac{\text{S}_{cjt}}{\sum_k \text{S}_{ckt}}$).

\[
\Omega_{ijt} = \sum_c \text{U}_{ict} \left(\frac{\text{S}_{cjt}}{\sum_k \text{S}_{ckt}}\right), \quad 1-\Phi_{jt} = \sum_c \left(\frac{\text{Imp}_{ct}}{\sum_k s_{ckt}}\right) \left(\frac{\text{S}_{cjt}}{\sum_k \text{S}_{ckt}}\right)
\]

Note that I assume a common import ratio for each good $j$ across all industries $i$. This is due to BEA data limitations, and corresponds with their construction of the import matrix in the Make and Use tables \citep{horowitzConceptsMethodsUS2009}. I define sectoral goods $j$ as tradeable if the average import ratio across all years in my sample is above 25\%. This yields 9 tradeable industries: ``Apparel and leather and allied products'', ``Computer and electronic products'', ``Electrical equipment, appliances, and components'', ``Furniture and related products'', ``Machinery'', ``Miscellaneous manufacturing'', ``Motor vehicles, bodies and trailers, and parts'', ``Oil and gas extraction'', and ``Textile mills and textile product mills''. I set $\Phi_{jt} = 1$ for all other industries.

From the GDP by Industry series, I use the Gross Output tables. These tables provide a ``Chain-Type Price Index for Gross Output'' measure by industry at the summary-level, which is prepared by combining the price indices of the commodities that the industry produces in a Fisher index-number formula. This provides an exact measure of price changes, presuming industry price is a stable flexible function of commodity prices \citep{diewertExactSuperlativeIndex1976}.

\begin{figure}
\centering 
\includegraphics[width=\textwidth]{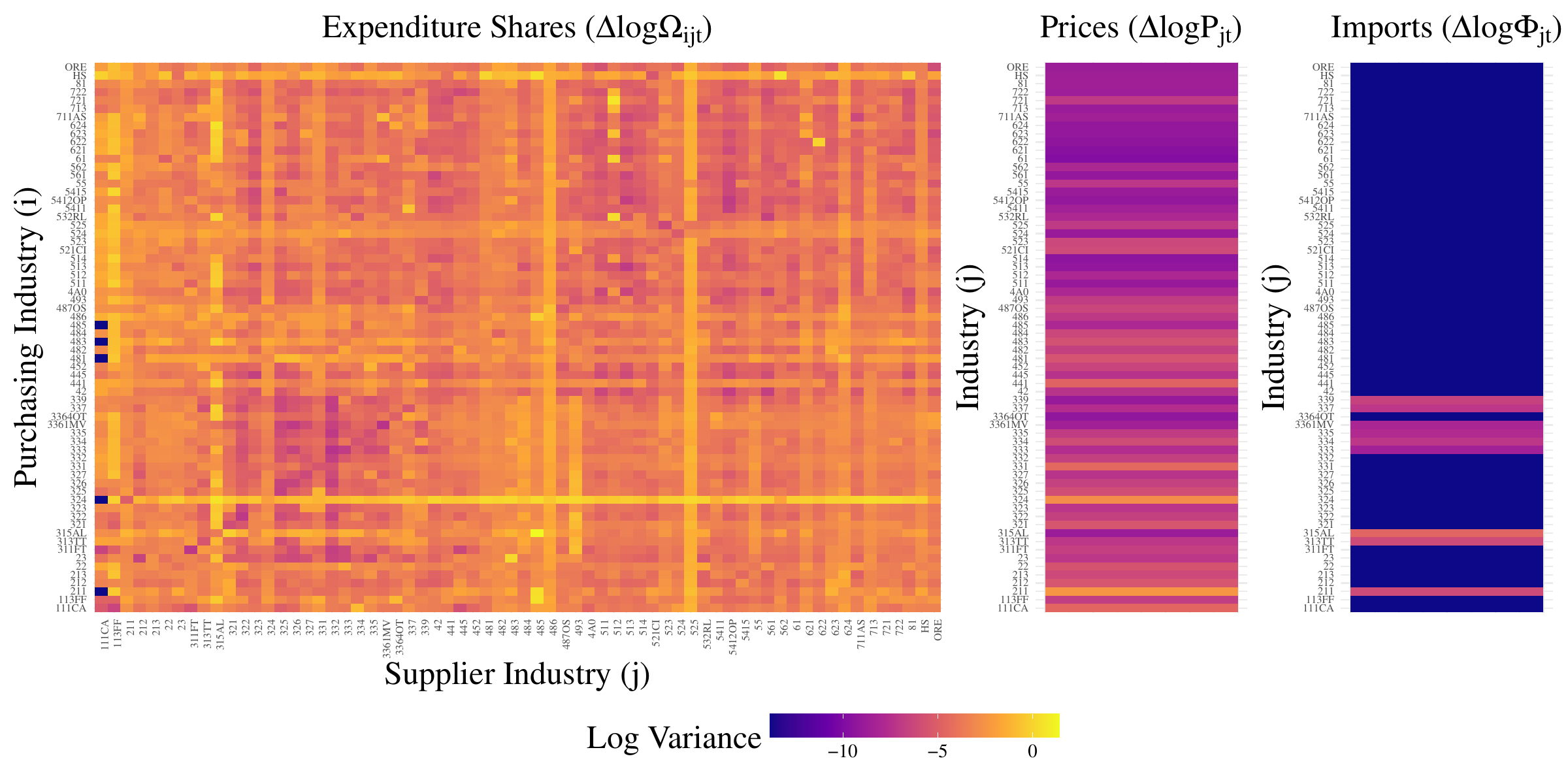}
\caption{Identifying Variation}
\label{fig:data_heatmap}
\end{figure}

I take logs and compute year-over-year changes in all three constructed variables. Unfiltered, this yields a balanced panel of 66 industries, 66 inputs, and 27 years (1998-2024), for a total of 117,612 observations; this variation is visualized in Figure \ref{fig:data_heatmap}. However, to ensure my estimates are driven by meaningful supplier-customer relationships, I drop industry-input pairs where the average expenditure share across all years in my sample is below 1\%. This leaves me with 31,455 observations.

\paragraph{Estimation} Since my empirical specification in Equation \ref{eq:reg_spec} is non-linear in substitution elasticities, I cannot recover my parameters of interest via linear regression. Instead, I estimate sector-specific intermediate input substitution elasticities $\hat \theta_i$ and Armington substitution elasticity $\hat \xi$ using generalized method of moments (GMM). I construct moments corresponding to the sector-specific coefficients on prices and import ratios implied by Equation \ref{eq:reg_spec}.
\[
\mathbb E[\mathbb I[i=i] \times \Delta \log P_{jt} \times \epsilon_{ijt}] = 0; \quad \mathbb E[\mathbb I[i=i] \times \Delta \log \Phi_{jt} \times \epsilon_{ijt}] = 0 
\]
This gives me $2N$ moments for $N+1$ parameters, which I weight evenly. I estimate $\hat \theta_i$ and $\hat \xi$ by minimizing the GMM objective function, and compute heteroskedasticity-robust standard errors using the standard GMM formula. I also estimate biased intermediate input substitution elasticities $\hat \theta_i^{\text{B}}$ ignoring imports, and a uniform intermediate input substitution elasticity $\bar \theta$ using the corresponding aggregate moments. More details on my estimation procedure can be found in Appendix \ref{ap:empirical:estimation}.

\subsection{Empirical Results}
\label{subsec:empirical_results}

\begin{figure}[!h]
    \centering 
    \includegraphics[width=.9\textwidth]{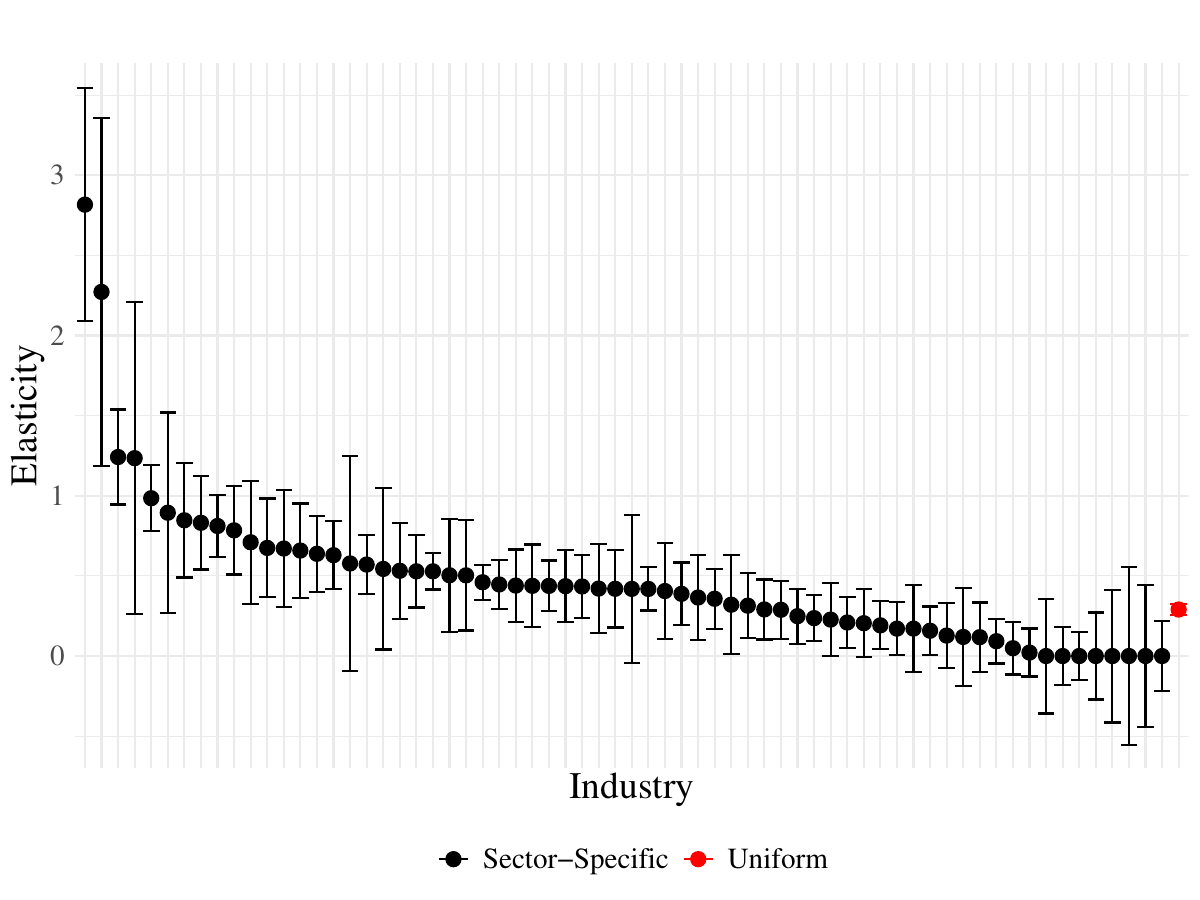}
    \caption{Uniform vs. Sector-Specific Intermediate Input Substitution Elasticities}
    \label{fig:elasticity_1}
\end{figure}

\begin{figure}[!h]
    \centering 
    \includegraphics[width=.9\textwidth]{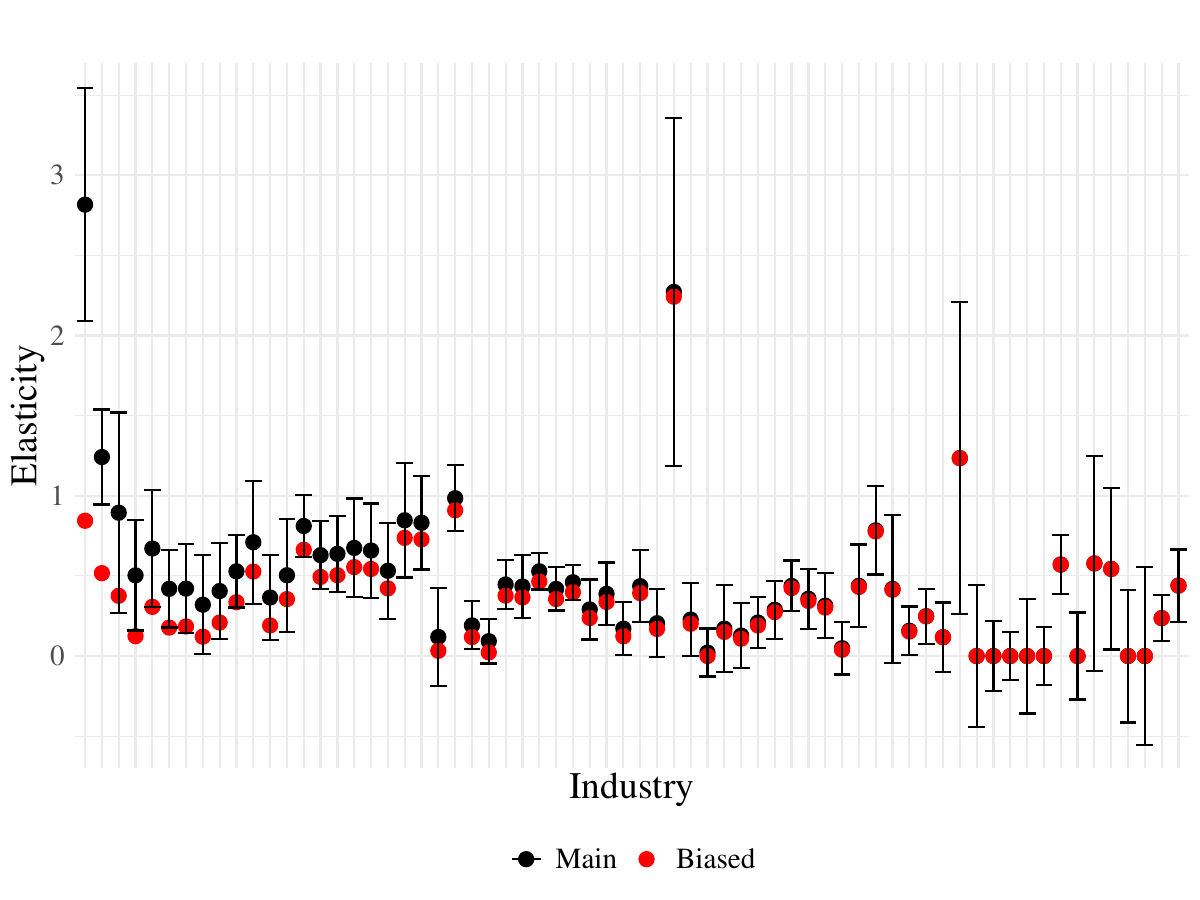}
    \caption{Import Bias of Intermediate Input Substitution Elasticities}
    \label{fig:elasticity_2}
\end{figure}

Sector-specific intermediate input substitution elasticities $\hat \theta_i$ and Armington substitution elasticity $\hat \xi$ are presented in Table \ref{tab:theta_estimates} and summarized in Figures \ref{fig:elasticity_1} and \ref{fig:elasticity_2}.

8 industries have estimated elasticities of 0.0, implying Leontief production of their intermediate input bundles: ``Rail transportation'', ``Water transportation'', ``Other real estate'', ``Forestry, fishing, and related activities'', ``Paper products'', ``Food and beverage and tobacco products'', ``Oil and gas extraction'', and ``Petroleum and coal products''. These sectors are mostly in manufacturing or depend on oil and gas as their primary intermediate input (e.g. transportation). 

\scriptsize{\begin{longtable}{llcccc}
\caption{Elasticity Estimates by Industry} \label{tab:theta_estimates} \\
\toprule
Industry & Code & Est. & SE & Biased Est. & Biased SE \\
\midrule
\endfirsthead
\multicolumn{6}{c}%
{{\bfseries \tablename\ \thetable{} -- continued from previous page}} \\
\toprule
Industry & Code & Est. & SE & Biased Est. & Biased SE \\
\midrule
\endhead
\midrule \multicolumn{6}{r}{{Continued on next page}} \\
\endfoot
\bottomrule
\endlastfoot
Rail transportation & 482 & 0.000 & 0.132 & 0.000 & 0.132 \\
Other real estate & ORE & 0.000 & 0.268 & 0.000 & 0.268 \\
Water transportation & 483 & 0.000 & 0.336 & 0.000 & 0.341 \\
Petroleum and coal products & 324 & 0.000 & 0.251 & 0.000 & 0.255 \\
Forestry, fishing, and related activities & 113FF & 0.000 & 0.165 & 0.000 & 0.166 \\
Oil and gas extraction & 211 & 0.000 & 0.091 & 0.000 & 0.105 \\
Paper products & 322 & 0.000 & 0.109 & 0.000 & 0.111 \\
Food and beverage and tobacco products & 311FT & 0.000 & 0.217 & 0.000 & 0.217 \\
Plastics and rubber products & 326 & 0.022 & 0.091 & 0.000 & 0.088 \\
Hospitals & 622 & 0.049 & 0.100 & 0.039 & 0.100 \\
Nonmetallic mineral products & 327 & 0.093 & 0.085 & 0.023 & 0.080 \\
Air transportation & 481 & 0.118 & 0.132 & 0.117 & 0.123 \\
Waste management and remediation services & 562 & 0.119 & 0.185 & 0.033 & 0.186 \\
Accommodation & 721 & 0.128 & 0.124 & 0.108 & 0.125 \\
Amusements, gambling, and recreation industries & 713 & 0.158 & 0.092 & 0.153 & 0.092 \\
Warehousing and storage & 493 & 0.171 & 0.165 & 0.150 & 0.167 \\
Chemical products & 325 & 0.171 & 0.100 & 0.123 & 0.115 \\
Other services, except government & 81 & 0.191 & 0.091 & 0.118 & 0.089 \\
Nursing and residential care facilities & 623 & 0.205 & 0.128 & 0.169 & 0.129 \\
Mining, except oil and gas & 212 & 0.209 & 0.097 & 0.191 & 0.103 \\
Food and beverage stores & 445 & 0.228 & 0.138 & 0.202 & 0.140 \\
Food services and drinking places & 722 & 0.237 & 0.086 & 0.237 & 0.086 \\
Utilities & 22 & 0.249 & 0.104 & 0.247 & 0.103 \\
Rental and leasing services and lessors of inta... & 532RL & 0.289 & 0.110 & 0.274 & 0.111 \\
Motor vehicle and parts dealers & 441 & 0.291 & 0.114 & 0.237 & 0.113 \\
Truck transportation & 484 & 0.314 & 0.122 & 0.303 & 0.115 \\
Fabricated metal products & 332 & 0.321 & 0.187 & 0.121 & 0.194 \\
Farms & 111CA & 0.357 & 0.113 & 0.344 & 0.114 \\
Computer and electronic products & 334 & 0.365 & 0.160 & 0.191 & 0.171 \\
Administrative and support services & 561 & 0.388 & 0.118 & 0.337 & 0.119 \\
Other transportation equipment & 3364OT & 0.406 & 0.181 & 0.209 & 0.203 \\
Primary metals & 331 & 0.419 & 0.081 & 0.356 & 0.084 \\
Pipeline transportation & 486 & 0.419 & 0.280 & 0.413 & 0.282 \\
Furniture and related products & 337 & 0.420 & 0.147 & 0.178 & 0.154 \\
Machinery & 333 & 0.421 & 0.168 & 0.185 & 0.176 \\
Electrical equipment, appliances, and components & 335 & 0.434 & 0.119 & 0.366 & 0.124 \\
Educational services & 61 & 0.436 & 0.137 & 0.393 & 0.140 \\
Support activities for mining & 213 & 0.438 & 0.096 & 0.424 & 0.109 \\
Other transportation and support activities & 487OS & 0.438 & 0.156 & 0.431 & 0.157 \\
Federal Reserve banks, credit intermediation, a... & 521CI & 0.440 & 0.137 & 0.440 & 0.137 \\
Wood products & 321 & 0.447 & 0.092 & 0.377 & 0.092 \\
Construction & 23 & 0.460 & 0.066 & 0.398 & 0.070 \\
Textile mills and textile product mills & 313TT & 0.504 & 0.209 & 0.124 & 0.161 \\
Legal services & 5411 & 0.504 & 0.213 & 0.355 & 0.242 \\
Miscellaneous manufacturing & 339 & 0.529 & 0.068 & 0.465 & 0.072 \\
Motor vehicles, bodies and trailers, and parts & 3361MV & 0.529 & 0.136 & 0.336 & 0.143 \\
Data processing, internet publishing, and other... & 514 & 0.532 & 0.181 & 0.422 & 0.203 \\
Performing arts, spectator sports, museums, and... & 711AS & 0.544 & 0.305 & 0.544 & 0.305 \\
Securities, commodity contracts, and investments & 523 & 0.571 & 0.111 & 0.571 & 0.111 \\
Funds, trusts, and other financial vehicles & 525 & 0.578 & 0.406 & 0.578 & 0.406 \\
Other retail & 4A0 & 0.630 & 0.128 & 0.494 & 0.143 \\
Wholesale trade & 42 & 0.638 & 0.144 & 0.504 & 0.158 \\
Miscellaneous professional, scientific, and tec... & 5412OP & 0.658 & 0.178 & 0.543 & 0.201 \\
Printing and related support activities & 323 & 0.671 & 0.222 & 0.307 & 0.137 \\
Management of companies and enterprises & 55 & 0.675 & 0.187 & 0.555 & 0.214 \\
Motion picture and sound recording industries & 512 & 0.710 & 0.232 & 0.527 & 0.296 \\
Transit and ground passenger transportation & 485 & 0.784 & 0.167 & 0.777 & 0.168 \\
Ambulatory health care services & 621 & 0.812 & 0.117 & 0.664 & 0.146 \\
Broadcasting and telecommunications & 513 & 0.832 & 0.177 & 0.728 & 0.204 \\
Computer systems design and related services & 5415 & 0.847 & 0.216 & 0.738 & 0.249 \\
General merchandise stores & 452 & 0.894 & 0.379 & 0.376 & 0.263 \\
Publishing industries, except internet (include... & 511 & 0.985 & 0.125 & 0.909 & 0.149 \\
Insurance carriers and related activities & 524 & 1.235 & 0.591 & 1.235 & 0.591 \\
Apparel and leather and allied products & 315AL & 1.242 & 0.180 & 0.517 & 0.662 \\
Housing & HS & 2.272 & 0.657 & 2.242 & 0.741 \\
Social assistance & 624 & 2.817 & 0.441 & 0.844 & 0.335 \\
\midrule
Uniform & --- & 0.290 & 0.022 & --- & --- \\
Armington & --- & 1.448 & 0.184 & --- & --- \\
\end{longtable}
}
\normalsize

The bulk of sectors have estimated intermediate input substitution elasticities $\in (0,1)$, implying gross complementarity across intermediate inputs. Manufacturing and mining industries (Codes $2-$ and $3-$) mostly have lower elasticities, while trade and information industries (Codes $4-$ and $5-$) mostly have higher elasticities. 4 industries have estimated elasticities above 1, implying gross substitutability across intermediate inputs: ``Social assistance'', ``Housing'', ``Insurance carriers and related activities'', and ``Apparel and leather and allied products''. The first three are industries whose gross output likely does not map neatly onto a production function, e.g. intermediate input spending is a small fraction of gross output in ``Housing'', which is mostly imputed rent of owner-occupied housing. The last is a manufacturing industry that is mostly offshored: the United States imports $\approx 85\%$ of apparel and leather allied products used domestically.

The uniform intermediate input substitution elasticity estimate is $\bar \theta = 0.29$, and is plotted in Figure \ref{fig:elasticity_1} alongside the sector-specific estimates. It is lower than most sector-specific estimates: 42 of the 66 industries have $\hat \theta_i > \bar \theta$. The Armington substitution elasticity estimate is $\hat \xi = 1.45$. 

Biased intermediate input substitution elasticities are shown in Table \ref{tab:theta_estimates} and summarized in Figure \ref{fig:elasticity_2}. They closely correspond with unbiased estimates for most industries. However, several sectors have meaningfully different estimates when ignoring imports; the largest discrepancies are in ``Social assistance'', ``Apparel and leather and allied products'', and ``General merchandise stores''. All three sectors have expenditure shares $>1\%$ on tradeable industries with variable import ratios (e.g. ``Apparel and leather allied products'', ``Textile mills and textile product mills'', and ``Motor vehicles, bodies and trailers, and parts''). Ignoring imports in these industries biases estimates of their intermediate input substitution elasticities upwards by $>0.5$.

\paragraph{Discussion} 

My estimates of intermediate input substitution elasticities can be summarized in three facts. (1) Intermediate inputs are gross complements in the vast majority of US industries: 62 of 66 industries have estimated elasticities below 1, and 52 have 90\% confidence intervals below 1. (2) Ignoring imports in production is trivial for most industries but biases estimates upwards for several sectors that rely heavily on tradeable goods with variable import ratios. (3) The uniform intermediate input substitution elasticity is not representative of US industries' true production functions. 

(3) is my most important result, and warrants further discussion. As shown in Figure \ref{fig:elasticity_1}, the uniform estimate $\bar \theta$ masks meaningful heterogeneity across sectors in their ability to substitute intermediate inputs: 26/66 industries have 90\% confidence intervals that exclude $\bar \theta = .290$. Moreover, the uniform estimate is biased downwards relative to the median sector-specific estimate. This bias is a consequence of industries with low substitution elasticities having more variable industry-input expenditure shares: as visualized in Figure \ref{fig:data_heatmap}, ``Oil and gas extraction'' (211) and ``Petroleum and coal products'' (324) have the most variable expenditure shares of all industries, meaning they dominate the variation identifying $\bar \theta$. 

My estimates are broadly in line with \citet{poirierReallocationDynamicsProduction2025}. We both find gross complementarity across inputs in sectoral production: their elasticities are less than one in all but one sector. Their uniform elasticity estimate is 0.22, only slightly lower than my uniform estimate of 0.29. Since they define sectors at a more aggregate level, these estimates line up nicely: as goods are defined more broadly, they become more differentiated, and less able to substitute for one another.

However, there are some key discrepancies sector-by-sector. ``Petroleum and coal products'' has the lowest elasticity estimate in my sample, $\hat \theta_{P} = 0.0$, implying it is unable to substitute inputs in its production. In contrast, ``Petroleum and coal products'' has the highest elasticity of all 32 sectors in \citet{poirierReallocationDynamicsProduction2025}, $\hat \theta_P = 1.08$, implying the opposite. Similarly, they estimate a low elasticity for ``Finance, insurance, real estate, rental and leasing'' relative to other sectors, while my corresponding estimates are higher, both in levels and relative to other industries. Finally, their estimates for ``Apparel and leather and allied products'' and ``Retail trade'' correspond to my biased estimates ignoring imports, but not my main estimates accounting for variation in the import ratio. This is to be expected since they assume a closed economy. All these differences alter the implied macro effect of shocks to the suppliers of these sectors (e.g. shocks to crude oil production are implied to have larger macro effects in my estimates).

My estimate of the Armington substitution elasticity $\hat \xi = 1.45$ implies substitutability between domestic and imported varieties. It is lower than the uniform trade elasticity of 4.55 estimated by \citet{caliendoEstimatesTradeWelfare2015}. However, the tradeable industries in my sample are more narrowly defined, and several correspond to sectors \citet{caliendoEstimatesTradeWelfare2015} estimate to have especially low substitutability (e.g. they estimate elasticities of 1.52 for ``Machinery'', 1.01 for ``Auto''). Industry by industry confidence bands mostly overlap.

\section{Quantitative Results}
\label{sec:discussion}

As discussed in Section \ref{subsec:empirical_results}, my main empirical result is that the uniform intermediate input substitution elasticity used in prior work masks meaningful differences in sectors ability to substitute inputs and is biased downward relative to the median sector-specific elasticity. In this section, I quantify the implications of this result by estimating the general equilibrium defined in Section \ref{subsec:model}, calibrated to the US economy.

To calibrate my model, I normalize prices and wages with respect to a base year, where the economy is assumed to be at the un-shocked steady state. In all exercises, the base year is 2024. I define exogenous parameters as described in Table \ref{tab:calibration}. For the household elasticity of substitution $\nu$, I estimate Equation \ref{eq:reg_spec} using household expenditure shares constructed from the BEA Supply-Use tables. My estimate $\hat \nu = 0.568$ is comfortingly in line with existing work estimating the household elasticity of substitution across broad categories of goods \citep{poirierReallocationDynamicsProduction2025, cominStructuralChangeLongRun2021}. For a deeper discussion of my calibration procedure, see Appendix \ref{ap:quantitative}.

\begin{table}[h]
    \centering
    \begin{threeparttable}
    \caption{Calibration of exogenous parameters}
    \begin{tabular}{|c|l|}
    \hline
    \textbf{Parameter} & \textbf{Calibration} \\ \hline
    $\gamma_i$ & Labor compensation expenditure share at base year \\ \hline
    $\omega_{ij}$ & Intermediate input expenditure share at base year \\ \hline
    $\beta_i$ & Personal consumption expenditure share at base year \\ \hline
    $L_i$ & Via market clearing \tnote{1} \\ \hline
    ${\sigma}$ & Set to .6 \tnote{2} \\ \hline
    $\nu, \theta_i, \xi$ & Estimated using Equation \ref{eq:reg_spec} \\ \hline
    \end{tabular}
    \begin{tablenotes}
    \footnotesize
    \item[1] I assume labor allocations correspond to the closed economy steady state.
    \item[2] See \citet{oberfieldMicroDataMacro2021} and \citet{carvalhoSupplyChainDisruptions2021}
    \end{tablenotes}
    \label{tab:calibration}
    \end{threeparttable}
\end{table}

In all my exercises, I compare the response of equilibrium values under two elasticity calibrations: (1) sector-specific intermediate input substitution elasticities $\hat \theta_i$, and (2) the uniform intermediate input substitution elasticity $\bar \theta$. This allows me to isolate the role of sector-specific substitution in changing the predicted effect of exogenous shocks. In my first set of exercises, I examine how sector-specific substitution mediates the \emph{propagation} of foreign price shocks and sectoral productivity shocks to prices. In my second set of exercises, I examine how sector-specific substitution mediates the \emph{aggregation} of sectoral productivity shocks to GDP. In Appendix \ref{ap:quantitative}, I repeat all exercises assuming a closed economy, and find results are similar.

\subsection{Shock Propagation to Prices with Sector-Specific Substitution} 
\label{subsec:propagation}

First, I simulate foreign price shocks as exogenous increases in the prices of imported sectoral goods, $\tilde P_{jt}$. These shocks are isomorphic to productivity shocks to foreign producers or import tariffs.\footnote{Since I only examine prices, production functions are CRS, and sectoral labor supply is exogenously fixed, redistributing tariff revenue is irrelevant.} I shock each tradeable sector $j$ one at a time, increasing its import price by 25\%, and store the response of domestic sectoral prices under my two elasticity calibrations.

\begin{table}[!h]
    \centering 
    \caption{Response of Sectoral Prices to Foreign Price Shocks}
    \label{tab:foreign_price_shocks}
    \begin{tabular}{llll}
\toprule
Sector & Main & Uniform & Difference \\
\midrule
\midrule
\multicolumn{4}{l}{\textbf{Shock to Oil and gas extraction import prices}} \\
\midrule
Petroleum and coal products & 9.570\% & 9.372\% & 0.198\% \\
Support activities for mining & 5.381\% & 4.224\% & 1.157\% \\
Oil and gas extraction & 4.039\% & 3.798\% & 0.241\% \\
\midrule
\multicolumn{4}{l}{\textbf{Shock to Computer and electronic products import prices}} \\
\midrule
Computer and electronic products & 5.443\% & 7.087\% & -1.644\% \\
Other transportation equipment & 1.937\% & 2.070\% & -0.134\% \\
Motor vehicles, bodies and trailers, and parts & 1.693\% & 2.428\% & -0.734\% \\
\bottomrule
\end{tabular} \\
    \raggedright{\small{Table lists the three largest \% increases in US industries prices after a 25\% increase in the specified foreign price, using the ``Main'' sector-specific elasticities $\hat \theta_i$ and the ``Uniform'' elasticity $\bar \theta$.}}
\end{table}

In Table \ref{tab:foreign_price_shocks}, I report the largest 3 domestic price responses to foreign price shocks in two industries of particular interest: ``Oil and gas extraction'', e.g. imports of crude oil, and ``Computer and electronic products'', e.g. imports of semiconductors. As expected, prices in related industries rise the most under both elasticity calibrations, but \emph{the magnitudes of price responses differ.} In particular, imposing the uniform intermediate input substitution elasticity $\bar \theta$ across all sectors overestimates the propagation of import price shocks for semiconductors and underestimates the propagation of import price shocks for crude oil. This is because industries that depend on crude oil are uniquely unable to substitute intermediate inputs: I point estimate most of their elasticities at 0.0. While manufacturing industries that depend on semiconductors also exhibit gross complementarity across inputs, they are able to substitute more than the uniform elasticity implies since, as discussed above, $\bar \theta$ is biased downwards due to the high variance of expenditure shares in oil-related industries.

The differences in the magnitude of price responses across elasticity calibrations are non-trivial. The uniform elasticity calibration overestimates the price response of ``Motor vehicles, bodies and trailers, and parts'' to a semiconductor price shock by 0.74\%, meaning sector-specific substitution lowers the impact by an order of magnitude of 30.2\%. In contrast, the uniform calibration underestimates the price response of ``Support activities for mining'' to a crude oil price shock by 1.12\%, meaning sector-specific substitution increases the impact by an order of magnitude of 27.4\%.

Next, I simulate sectoral productivity shocks, exogenous changes in sectoral productivities $Z_{it}$. I calibrate the shocks using the Industry TFP measure reported in the BLS KLEMS data: I estimate the variance of quadrennial sectoral TFP growth from 1997-2023, and draw 1000 shocks from the corresponding multivariate normal distributions with mean 0. Each shock vector can thus be interpreted as a \emph{sectoral business cycle}. I solve for the equilibrium corresponding to each sectoral business cycle under my two elasticity calibrations, and compute the average price response for each sector.

\begin{table}[!h]
    \centering
    \caption{Response of Sectoral Prices to Sectoral Productivity Shocks}
    \label{tab:businesscycle_price}
    \begin{tabular}{llll}
\toprule
Industry & Main & Uniform & Difference \\
\midrule
Support activities for mining & 11.62\% & 6.31\% & 5.31\% \\
Oil and gas extraction & 6.43\% & 5.09\% & 1.34\% \\
Petroleum and coal products & 4.57\% & 3.37\% & 1.20\% \\
\bottomrule
\end{tabular}

\end{table}

In Table \ref{tab:businesscycle_price}, I list the three sectors with the largest increase in average price response due to sector-specific substitution. As expected, all three are involved in oil and gas production. These sectors each have intermediate input substitution elasticities point estimated at 0.00, less than the uniform elasticity of 0.29. This means sector-specific substitution increases the asymmetry of their response to productivity shocks as described in Section \ref{subsec:stylized_example}: negative shocks generate larger price increases, while positive shocks generate smaller price decreases. These sectors are also interdependent on one another, which amplifies the asymmetry further. 

Again, the role of sector-specific substitution in the predicted effects of sectoral business cycles is non-trivial. As shown in Table \ref{tab:businesscycle_price}, the uniform elasticity calibration underestimates the average price response of ``Petroleum and coal products'' (e.g. gas prices) to a sectoral business cycle by 1.20\%, meaning sector-specific substitution increases the price response by an order of magnitude of 35.6\%.

\subsection{Shock Aggregation to GDP with Sector-Specific Substitution}

First, I generate I generate \emph{severe} productivity shocks to individual sectors as TFP shocks that cut sectoral productivity by 25\%. I shock each industry one by one, and solve for the equilibrium corresponding to each shock under my two elasticity calibrations. I store the response of GDP. 

\begin{table}[!h]
    \centering 
    \caption{Response of GDP to Severe Sectoral Productivity Shocks}
    \label{tab:severe_prod_shocks}
    \begin{tabular}{llll}
\toprule
Industry & Main & Uniform & Change \\
\midrule
\midrule
\multicolumn{2}{l}{\textbf{Smaller GDP loss}} \\
\midrule
Insurance carriers and related activities & -5.35\% & -7.83\% & 2.48\% \\
Federal Reserve banks, credit intermediation, and ... & -5.06\% & -5.72\% & 0.66\% \\
Construction & -1.47\% & -1.79\% & 0.32\% \\
\midrule
\multicolumn{2}{l}{\textbf{Larger GDP loss}} \\
\midrule
Chemical products & -9.62\% & -9.44\% & -0.18\% \\
Farms & -7.02\% & -6.75\% & -0.27\% \\
Paper products & -1.70\% & -1.40\% & -0.30\% \\
\bottomrule
\end{tabular} \\
    \raggedright{\small{This table lists the three industries largest \% increases and largest \% decreases in the GDP effect of a 25\% productivity shock using the ``Main'' sector-specific elasticities $\hat \theta_i$ versus the ``Uniform'' elasticity $\bar \theta$.}}
\end{table}

In Table \ref{tab:severe_prod_shocks}, I report the the three sectors where sector-specific substitution increases and decreases the GDP effect of a severe sectoral productivity shock the most. As expected, sector-specific substitution increases the GDP effect of shocks to sectors whose output is used in manufacturing industries with low substitution elasticities: ``Chemical products'' output is used in the production of ``Petroleum and coal products'', which has an elasticity estimate of 0.0; ``Farms'' output is used in the production of ``Food and beverage and tobacco products'', which has an elasticity estimate of 0.0; and ``Paper products'' output is used in the production of ``Plastic and rubber products'' which has an elasticity estimate of 0.02. In contrast, sector-specific substitution decreases the GDP effect of shocks to sectors whose output is used in industries with high substitution elasticities: the outputs of ``Insurance carriers'', ``Federal Reserve banks'', and ``Construction'' are all used mostly by the ``Housing'' sector, which has an elasticity estimate of 2.27 (as discussed in Section \ref{subsec:empirical_results}, this is likely due to the fact that gross output in ``Housing'' doesn't map well onto a production function). The largest increase in GDP loss due to sector-specific substitution is .30\% for ``Paper products'', an increase in magnitude of 21.4\%. The largest decrease in GDP loss is 2.48\% for ``Insurance carriers'', a decrease in magnitude of 31.7\%.

Next I simulate sectoral business cycles as in Section \ref{subsec:propagation}: I draw 1000 shock vectors from a multivariate normal distribution calibrated to match the variance of quadrennial sectoral TFP growth from 1997-2023. I solve for the equilibrium corresponding to each shock under my two elasticity calibrations and a Cobb-Douglas calibration where $\theta_i = 1$ for all $i$. 

\begin{table}[!h]
    \centering 
    \caption{Response of GDP to Sectoral Business Cycles}
    \label{tab:gdp_business_cycles}
    \begin{tabular}{llll}
\toprule
Calibration & Mean & Std Dev & Skewness \\
\midrule
Main & -1.63\% & 3.89\% & 0.04\% \\
Uniform & -1.98\% & 3.91\% & -0.00\% \\
\bottomrule
\end{tabular}
 \\
    \raggedright{\small{This table lists the summary statistics of the GDP response to simulated sectoral business cycles, using the ``Main'' sector-specific elasticities $\hat \theta_i$, and the ``Uniform'' elasticity $\bar \theta$.}}
\end{table}

\begin{figure}[!h]
    \centering 
    \includegraphics[width=.9\textwidth]{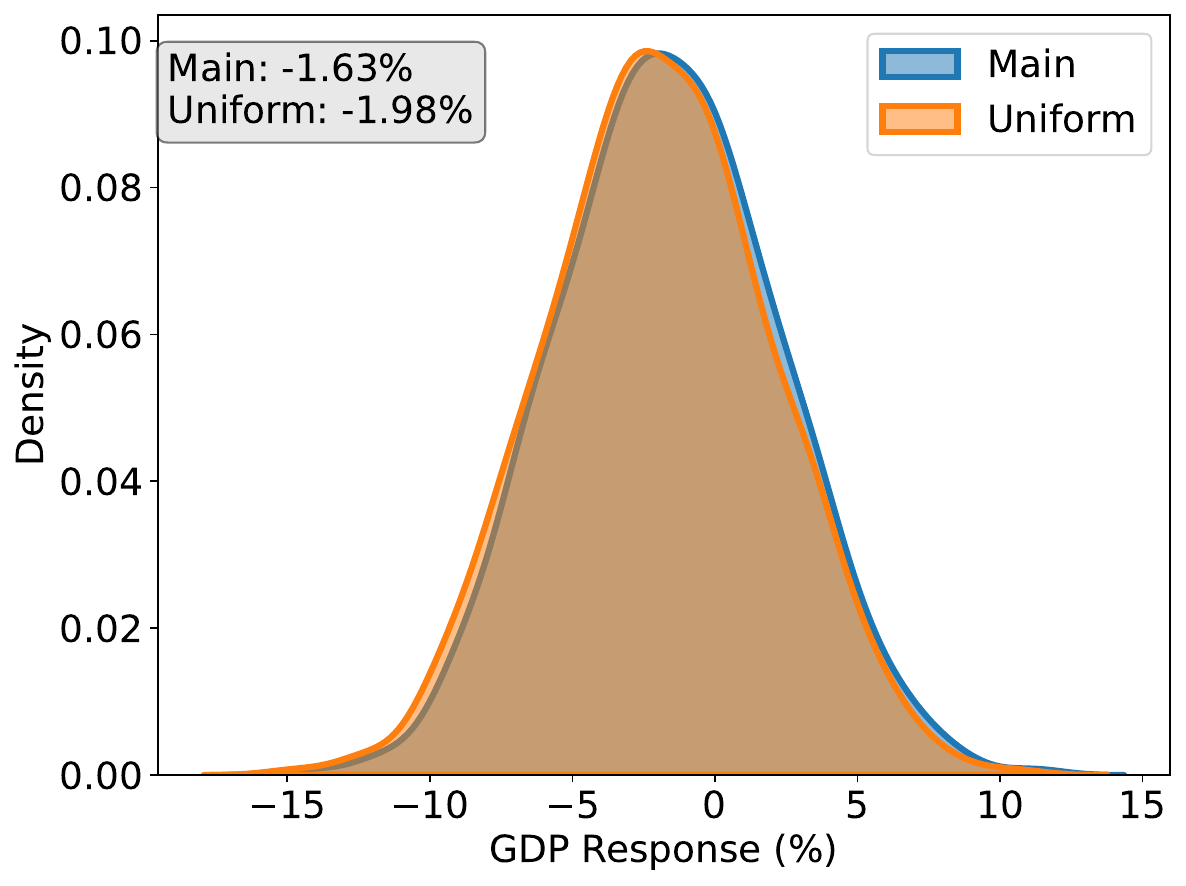}
    \caption{Response of GDP to Sectoral Business Cycles}
    \label{fig:gdp_business_cycles}
    \raggedright{\small{This figure plots the distribution of GDP response to sectoral business cycles, using the ``Main'' sector-specific elasticities $\hat \theta_i$, and the ``Uniform'' elasticity $\bar \theta$. The text box reports the average response.}}
\end{figure}

The resulting GDP distributions are summarized in Table \ref{tab:gdp_business_cycles} and plotted in Figure \ref{fig:gdp_business_cycles}. Sector-specific substitution decreases the standard deviation of GDP fluctuations by 0.02\% and increases the skew by 0.04\%. Most notably, it increases the mean of the GDP distribution by 0.35\%. This is due to the downward bias of the uniform elasticity relative to the median sector-specific elasticity, discussed in Section \ref{subsec:empirical_results}: most sectors are more able to substitute inputs than the uniform elasticity implies, which dampens the asymmetric response of GDP to sectoral shocks described in Section \ref{subsec:stylized_example}.

This change in the mean GDP response is non-trivial. It means sector-specific substitution decreases the average GDP loss from a sectoral business cycle by 0.35\%, implying such business cycles are 17.7\% less costly than predicted when ignoring differences in industries ability to substitute inputs. This result can be seen as an extension of the welfare cost of business cycles described in \citet{lucasModelsBusinessCycles1987}: here the cost is driven by non-linearities in production as well as in utility, and sector-specific substitution lowers the cost because production is less non-linear in most sectors than the uniform elasticity estimate implies.

\section*{Conclusion}

Production function parameters governing the substitutability of inputs play a central role in determining the effect of sectoral shocks on prices and GDP. In this paper I provide new estimates of these parameters for US industries, permitting them to vary across sectors, by using a novel empirical strategy that relies on the unique within-industry/across-input variation provided by the BEA Input-Output Accounts. My estimates document new facts about US sectoral production, most notably that the ability of US industries to substitute inputs varies meaningfully across sectors, making the uniform estimate of the intermediate input substitution elasticity unrepresentative and biased downwards relative to the median sector-specific estimate.

I quantify the macroeconomic implications of this result using a GE model of multi-sector production calibrated to the US economy. I find that ignoring sector-specific substitution changes the predicted effect of import price shocks and sectoral productivity shocks on prices in several key industries. Most notably, oil and gas related industries' have larger price responses to import shocks and sectoral business cycles due to their inability to substitute inputs.

Similarly, I find that ignoring sector-specific substitution changes the predicted effect of sectoral productivity shocks on GDP. Severe shocks to industries that supply manufacturing sectors (Farms, Chemical products, and Paper products) have larger GDP effects due to the inability of their customers to substitute inputs. In contrast, severe shocks to industries that primarily supply the Housing sector (Insurance carriers, Federal Reserve banks, and Construction) have smaller GDP effects. Finally, I find that sector-specific substitution decreases the average GDP loss from sectoral business cycles by 0.35\%, implying such business cycles are 17.7\% less costly than predicted when ignoring differences in industries ability to substitute inputs.

While I focus on shocks to sectoral TFP and import prices, my results have more general implications. The intermediate input substitution elasticity determines how sectors respond to changes in the prices of inputs. Thus the sector-to-sector propagation of any phenomena that manifests in prices will be affected by heterogeneity in this parameter, including distortions \citep{bigioDistortionsProductionNetworks2020} and price rigidities \citep{rubboNetworksPhillipsCurves2023} \citep{laoOptimalMonetaryPolicy2022}. I leave these implications for future research.

\pagebreak 


\bibliography{../../../mylibrary.bib} 
\clearpage

\appendix
\numberwithin{equation}{section}
\setcounter{section}{0} \renewcommand{\thesection}{\Alph{section}}
\setcounter{equation}{0} \renewcommand{\theequation}{\Alph{section}.%
\arabic{equation}}
\setcounter{table}{0} \renewcommand{\thetable}{\Alph{section}.\arabic{table}}%
\setcounter{figure}{0} \renewcommand{\thefigure}{\Alph{section}.\arabic{figure}}%

\section{Theoretical Appendix}
\label{ap:theory}

Note that the Proofs of Theorems \ref{thm:sales_share}, \ref{thm:prices} and \ref{thm:gdp} closely follow the derivation in \citet{baqaeeMicroPropagationMacro2023} and \citet{baqaeeMacroeconomicImpactMicroeconomic2019}.

\paragraph{Proof of Theorem \ref{thm:sales_share}}
Setting aggregate expenditure as the numeraire, $\sum_i P_{it} C_{it} = 1$, market clearing delivers the following expression relating Domar weights, the input-output matrix, and export expenditures.
\begin{align*}
Y_{jt} = \sum_i X_{ijt} + C_{jt}^f \rightarrow \lambda_{jt} = \sum_i a_{ijt} \lambda_{it} + P_{jt} C_{jt}^f \\
\boldsymbol \lambda_t = \boldsymbol \lambda'_t \mathbf A_t + \mathbf{NX}_t
\end{align*}
Taking the total derivative, we can express changes in Domar weights in terms of changes in export expenditures and changes in the input-output matrix.
\[
\partial \boldsymbol \lambda_t = (\boldsymbol \lambda^T_t \partial \mathbf A_t + \partial \mathbf{NX}_t) \boldsymbol \Psi_t \rightarrow \partial \lambda_{it} = \sum_j \psi_{jit} \left(\sum_k \lambda_{kt} \partial a_{kjt} + \partial NX_{jt}\right)
\]
By definition of export demand $C^f_{it} = \phi^f_i \left(\frac{P_{it}}{E_t}\right)^{-\tilde \xi}$, changes in export expenditures can be expressed in terms of current export expenditures and changes in prices.
\[
\partial NX_{jt} = NX_{jt} \partial \log NX_{jt} = NX_{jt} ((1-\tilde \xi) \partial \log P_{jt} + \partial \log E_t)
\]
And by cost-minimization, changes in the input-output matrix can be expressed in terms of current expenditure shares and changes in prices.
\[
\partial a_{ijt} = a_{ijt} \partial \log a_{ijt} = a_{ijt}\left((\theta_i - \sigma) \partial \log Q_{it} + (\xi-\theta_i) \log \bar P_{jt} + (1-\xi) \log P_{jt}\right)
\]
Cost-minimization also allows us to write changes in CES price indices in terms of changes in input prices and expenditure shares.
\[
\partial \log Q_{it} = \sum_j \Omega_{ijt} \partial \log \bar P_{jt}, \space \partial \log \bar P_{jt} = \Phi_{ijt}  \partial \log P_{jt} + (1 - \Phi_{ijt}) (\partial \log \tilde P_{jt} + \partial \log E_t)
\]
\paragraph{Proof of Theorem \ref{thm:prices}}
By cost-minimization of the firm problem, we can write changes in prices as follows:
\begin{align*}
\partial \log P_{it} &= \Gamma_{it} \partial \log W_{it} + (1-\Gamma_{it}) \partial \log Q_{it} - \partial \log Z_{it} \\
\partial \log Q_{it} &= \sum_j \Omega_{ijt} \partial \log \bar P_{jt} \\
\partial \log \bar P_{jt} &= \Phi_{ijt}  \partial \log P_{jt} + (1 - \Phi_{ijt}) (\partial \log \tilde P_{jt} + \partial \log E_t)
\end{align*}
Combining these equations, we get the forward propagation equations of price determination.
\begin{align*}
\partial \log P_{it} &= \Gamma_{it} \partial \log W_{it} + (1-\Gamma_i)\sum_j \Omega_{ijt} (\Phi_{ijt} \partial \log p_{jt} + (1-\Delta_{ij})\partial \log E_t \tilde P_{jt}) - \partial \log Z_{it} \\
&= \Gamma_{it} \partial \log W_{it} + \sum_j (a_{ijt} \partial \log P_{jt} + \tilde a_{ijt} (\partial \log E_t \tilde P_{jt}) - \partial \log Z_{it} \\
\partial \log \mathbf P_t &= \boldsymbol {\Gamma}_t \partial \log \mathbf W_t + \mathbf A_t \partial \log \mathbf P_t + \mathbf {\tilde A}_t (\partial \log \mathbf {\tilde P}_t + \partial \log E_t) - \partial \log \mathbf Z_t \\
\partial \log \mathbf P_t &= \mathbf \Psi_t \left(\boldsymbol {\Gamma}_t \partial \log \mathbf W_t + \mathbf {\tilde A}_t (\partial \log \mathbf {\tilde P}_t + \partial \log E_t) - \partial \log \mathbf Z_t\right)
\end{align*}
Foreign prices $\mathbf {\tilde P}_t$ and productivities $\mathbf Z_t$ are exogenous. Sector-specific wages (factor prices) are equal to the marginal revenue of labor, i.e.
\begin{align*}
W_{it} &= P_{it} Z_{it}^{1-\frac 1 \sigma} \left(\gamma_i \frac {Y_{it}}{L_{it}}\right)^{\frac 1 \sigma} \\
\partial \log W_{it} &= \partial \log P_{it} + \left(1-\frac 1 \sigma\right) \partial \log Z_{it} + \frac 1 \sigma \left(\partial \log Y_{it}\right)
\end{align*}
Observe that we can write changes in sectoral output in terms of changes in sales share (remember that we've set nominal expenditures as our numeraire).
\[
\lambda_{it} = P_{it} Y_{it} \rightarrow \partial \log Y_{it} = \partial \log\lambda_{it} - \partial \log P_{it}
\]
Plugging into the wage equation, we have
\[
\partial \log W_{it} = \left(1-\frac 1 \sigma\right) (\partial \log P_{it} + \partial \log Z_{it}) + \frac 1 \sigma \partial \log \lambda_{it}
\]
\paragraph{Proof of Theorem \ref{thm:gdp}}
Since our economy is efficient, the decentralized equilibrium corresponds to the solution to the following social planner problem.
\begin{align*}
&\max_{C_{it}, X_{ijt}} C_t \\
&\text{ subject to } Y_{it} = F_{it}( Z_{it}, L_{it}, X_{ijt}, \tilde X_{ijt}) = C_{it} + \sum_j X_{jit} + C_{it}^f\quad [\mu_{it}] \\
&\text{ and } L_{it} = \bar L_i \quad [\iota_{it}]
\end{align*}
The envelope theorem implies that
\[
\frac{\partial C_t}{\partial \log Z_{it}} = Z_{it} \frac{\partial C_t}{\partial Z_{it}} = \mu_{it} Y_{it}
\]
First order conditions further imply that
\[
\mu_{jt} = \mu_{it} \frac{\partial{F_{it}}}{\partial X_{ijt}} \text{ for every } i, j
\]
Which correspond with the firm first-order conditions in the decentralized equilibrium, implying that $\mu_{it} = P_{it}$. Thus
\[
\frac{\partial C_t}{\partial \log Z_{it}} = P_{it} Y_{it} = \lambda_{it}
\]
\section{Empirical Appendix}
\label{ap:empirical}

\subsection{Estimation Details}
\label{ap:empirical:estimation}

I estimate $\theta_i, \xi$ by generalized method of moments (GMM). I first residualize all variables with respect to industry-year fixed effects. I am left with the following equation relating residualized log changes in expenditure shares, prices, and import ratios:
\[
\Delta \log \Omega_{ijt}^{\eta_{it}} = (1 - \theta_i) \Delta \log P_{jt}^{\eta_{it}} + \frac{\xi - \theta_i}{\xi - 1} \Delta \log \Phi_{ijt}^{\eta_{it}} + \epsilon_{ijt}
\]
This equation corresponds to a linear regression with $N$ industry-specific coefficients on the interaction term between industry and price changes, and $N$ industry-specific coefficients on the interaction term between industry and import ratio changes:
\[
\Delta \log \Omega_{ijt}^{\eta_{it}} = \beta^1_i \Delta \log P_{jt}^{\eta_{it}} \mathbb I[i = i] + \beta^2_i \Delta \log \Phi_{ijt}^{\eta_{it}} \mathbb I[i = i] + \epsilon_{ijt}, \text{ where } \beta_i^1 = (1-\theta_i), \; \beta_i^2 = \frac{\xi - \theta_i}{\xi - 1}.
\]
To estimate $\theta_i$ and $\xi$ via GMM, I construct moments corresponding to these regression coefficients:
\[
g(\theta, \xi) = \left[\mathbb{E}\!\left[\Delta \log P_{jt}^{\eta_{it}} \times \mathbb{I}[i=I] \times \epsilon_{ijt}\right] = 0, \; \mathbb{E}\!\left[\Delta \log \Phi_{ijt} \times \mathbb{I}[i=I] \times \epsilon_{ijt}\right] = 0\right].
\]
I construct the sample analogs as expected, where $\bar{N}$ denotes the total number of observations:
\[
\bar{g}(\theta, \xi) = \left[\frac{1}{\bar{N}} \sum \Delta \log P_{jt}^{\eta_{it}} \times \mathbb{I}[i=I] \times \epsilon_{ijt}, \; \frac{1}{\bar{N}} \sum \Delta \log \Phi_{ijt} \times \mathbb{I}[i=I] \times \epsilon_{ijt}\right].
\]
I estimate $\hat{\theta}_i, \hat{\xi}$ to minimize the GMM objective function. I have $2N$ moments for $N+1$ parameters, so the model is over-identified. I weight each moment equally (i.e., I use an identity weighting matrix):
\[
\hat{\theta}, \hat{\xi} = \arg \min_{\theta, \xi} Q_n(\theta, \xi) = \bar{g}(\theta, \xi)' \mathbf{I} \, \bar{g}(\theta, \xi).
\]
This is equivalent to minimizing the sum of squared moments. Under identity weighting, the asymptotic normality theorem for GMM implies that
\begin{align*}
\widehat{\text{Var}}(\hat{\theta}, \hat{\xi}) &= \frac{1}{N} (G' G)^{-1} G' \hat{\Omega} \, G (G' G)^{-1}, \\
G &= \frac{\partial \bar{g}(\theta, \xi)}{\partial (\theta, \xi)}, \quad \hat{\Omega} = \frac{1}{N} \sum_{i=1}^{N} g(\hat{\theta}, \hat{\xi}) \, g'(\hat{\theta}, \hat{\xi}).
\end{align*}
I recover standard errors as $\text{SE} = \sqrt{\text{diag}\!\left(\widehat{\text{Var}}(\hat{\theta}, \hat{\xi})\right)}$. Because the objective function is highly non-linear in the parameters, I use the Powell optimization algorithm, which is derivative-free.

\subsection{Replicating \citet{atalayHowImportantAre2017} Empirical Strategy}
\label{ap:atalay}

\citet{atalayHowImportantAre2017} constructs an instrumental variables for price changes using year-over-year shifts in military spending, the revenue share of the general federal government defense sector (GFGD), and the row-sums of the Leontief inverse of the input-output ``customer'' matrix.\footnote{\citet{miranda-pintoFlexibilityFrictionsMultisector2022} use the same instrument. Formal definitions can be seen in Equations 14, 15 and 16 of \citet{atalayHowImportantAre2017}.} 

I construct the Atalay instruments, and estimate the elasticity of substitution across intermediate inputs using it. I compare this estimate with the uniform elasticity estimate corresponding to my empirical specification in Equation \ref{eq:reg_spec}, which addresses endogeneity with industry-year fixed effects rather than an instrumental variable. I include an estimate assuming a closed economy, i.e. dropping the coefficient on changes in the import ratio, to emphasize that this term does not drive the differences between the two empirical strategies.

The results are summarized in Table \ref{tab:atalay_comparison}. Comfortingly, confidence intervals overlap. However, the Atalay IV estimate is negative, and has much larger standard errors, while my estimate is more realistic (positive) and much more precise. This discrepancy is to be expected, since the Atalay instrument is a weak predictor of price changes: the adjusted $R^2$ in the first stage is 0.002. 

I take this as strong evidence that addressing endogeneity due to unobserved industry-year shocks with industry-year fixed effects rather than an instrument is preferable when estimating the intermediate input substitution elasticity, as opposed to jointly estimating the intermediate input substitution elasticity and the substitution elasticity between labor and the intermediate input bundle as in \citet{atalayHowImportantAre2017}. In particular, Table \ref{tab:atalay_comparison} demonstrates how my empirical specification provides more identifying variation and delivers more precise estimates, which is what allows me to estimate sector-specific elasticities.

\begin{table}
\centering 
\caption{Comparison with \citet{atalayHowImportantAre2017} Empirical Strategy}
\label{tab:atalay_comparison}

\begin{tabular}{l c c c}
\hline
 & Atalay IV & Uniform (Closed Economy) & Uniform \\
\hline
Elasticity & $-0.40$  & $0.33$   & $0.29$   \\
           & $(0.91)$ & $(0.02)$ & $(0.02)$ \\
\hline
Num. obs.  & $30290$  & $30290$  & $30290$  \\
\hline
\end{tabular}

\end{table}

\section{Quantitative Appendix}
\label{ap:quantitative}

\paragraph{Calibration: } I calibrate by picking a base year in my data (2024), and assuming it correspond to the non-stochastic equilibrium of the model, where all prices ($W_{it}, P_{it}, E_t, \tilde P_{it}$) and productivities equal 1. This is equivalent to normalizing GDP and price responses as relative to the base year.

Under this normalization, all share parameters equal their expenditure shares at the base year.
\[
\gamma_i = \frac{W_i L_i}{P_i Y_i}; \quad \omega_{ij} = \frac{P_{j} X_{ij}}{Q_i M_i}; \quad \psi_{ij} = \frac{P_{jd} X_{ijd}}{P_j X_{ij}}; \quad \beta_i = \frac{P_i C_i}{P C} 
\]
I can back out sectoral labor allocations $L_i$ from these expenditure shares, \emph{assuming they correspond to the closed economy} (i.e. in the absence of trade, labor allocations would be the same). 
\[
\begin{aligned}
L_i = \gamma_i Y_i \\
X_{ijd} = (1-\gamma_i) \omega_{ij} Y_i \\
Y_i = C_i + \sum_j X_{jid}  = \beta_i + \sum_j (1-\gamma_j) \omega_{ji} Y_j \\
\mathbb Y = \boldsymbol \beta + (1-\gamma) \boldsymbol \Omega \mathbb Y \\
\rightarrow \mathbb Y = (\mathbb I - (1-\gamma) \boldsymbol \Omega)^{-1} \boldsymbol \beta
\end{aligned}
\]

With share parameters and labor allocations fixed, I am left with the parameters that determine the foreign demand schedule for exports of tradeable goods: demand shifters $\psi_i^f$ and export demand elasticity $\tilde \xi$. I back out demand shifters $\psi_i^f$ via market clearing in the base year. I assume the export demand elasticity $\tilde \xi$ equals the Armington elasticity $\xi$, which I estimate in Section \ref{sec:empirical}. 

\paragraph{Closed Economy: } I also consider a closed economy version of the model to ensure my quantitative results aren't driven by assumptions made to calibrate foreign demand. The results of each quantitative exercise under the closed economy assumption are reported below (I omit foreign price shocks, since they are irrelevant in a closed economy). Results are unchanged, which is to be expected since trade shares are small in most industries in the US.

\begin{table}[!h]
    \centering
    \caption{Response of Sectoral Prices to Sectoral Productivity Shocks (Closed)}
    \begin{tabular}{llll}
\toprule
Industry & Main & Uniform & Difference \\
\midrule
Support activities for mining & 21.45\% & 10.62\% & 10.83\% \\
Oil and gas extraction & 12.54\% & 9.20\% & 3.33\% \\
Petroleum and coal products & 12.61\% & 9.33\% & 3.28\% \\
\bottomrule
\end{tabular}

\end{table}

\begin{table}[!h]
    \centering 
    \caption{Response of GDP to Severe Sectoral Productivity Shocks (Closed)}
    \begin{tabular}{llll}
\toprule
Industry & Main & Uniform & Change \\
\midrule
\midrule
\multicolumn{2}{l}{\textbf{Smaller GDP loss}} \\
\midrule
Insurance carriers and related activities & -5.46\% & -8.06\% & 2.60\% \\
Federal Reserve banks, credit intermediation, and ... & -5.10\% & -5.88\% & 0.77\% \\
Primary metals & -3.87\% & -4.22\% & 0.35\% \\
\midrule
\multicolumn{2}{l}{\textbf{Larger GDP loss}} \\
\midrule
Chemical products & -10.13\% & -9.88\% & -0.25\% \\
Farms & -7.01\% & -6.76\% & -0.25\% \\
Paper products & -1.83\% & -1.49\% & -0.35\% \\
\bottomrule
\end{tabular} \\
    \raggedright{\small{This table lists the three industries largest \% increases and largest \% decreases in the GDP effect of a 25\% productivity shock using the ``Main'' sector-specific elasticities $\hat \theta_i$ versus the ``Uniform'' elasticity $\bar \theta$.}}
\end{table}

\begin{table}[!h]
    \centering 
    \caption{Response of GDP to Sectoral Business Cycles (Closed)}
    \begin{tabular}{llll}
\toprule
Calibration & Mean & Std Dev & Skewness \\
\midrule
Main & -1.94\% & 4.13\% & -0.01\% \\
Uniform & -2.32\% & 4.15\% & -0.02\% \\
\bottomrule
\end{tabular}
 \\
    \raggedright{\small{This table lists the summary statistics of the GDP response to simulated sectoral business cycles, using the ``Main'' sector-specific elasticities $\hat \theta_i$, and the ``Uniform'' elasticity $\bar \theta$.}}
\end{table}

\begin{figure}[!h]
    \centering 
    \includegraphics[width=.9\textwidth]{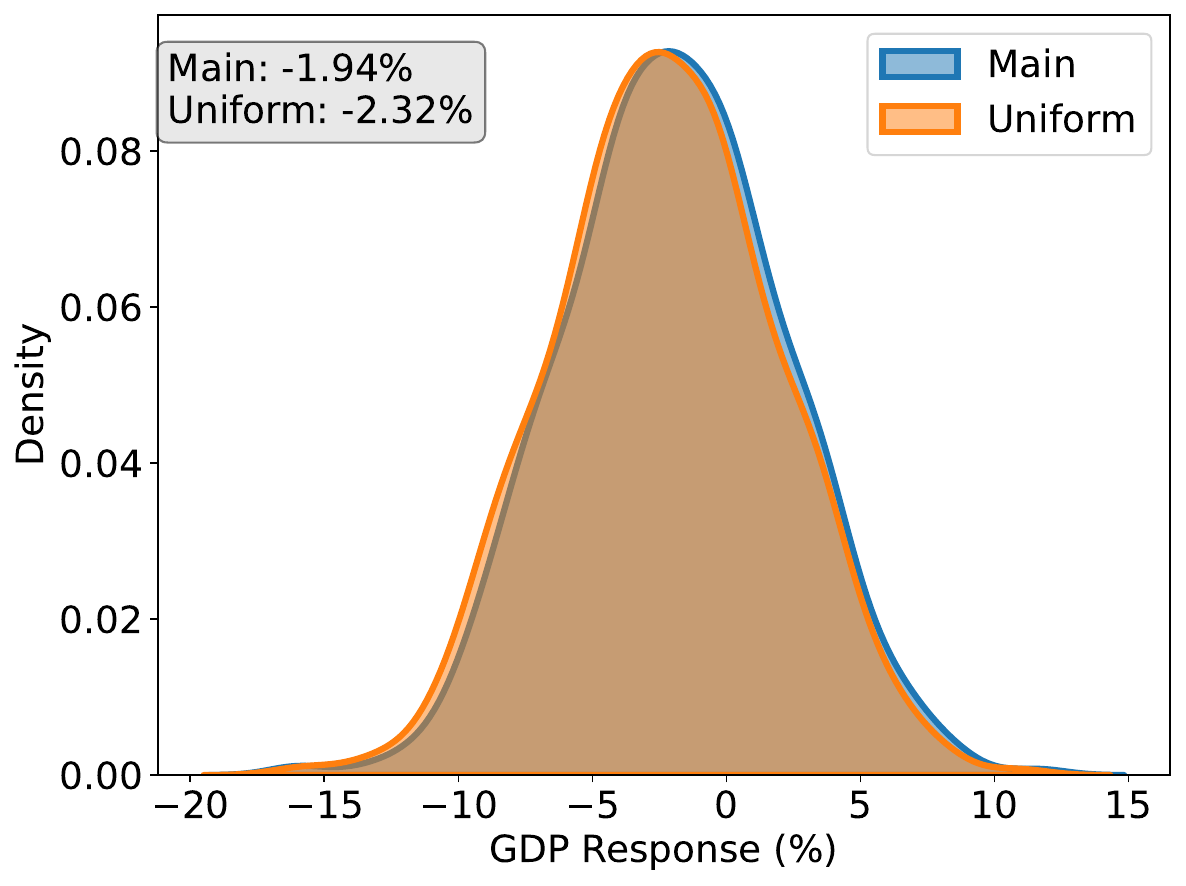}
    \caption{Response of GDP to Sectoral Business Cycles}
    \raggedright{\small{This figure plots the distribution of GDP response to sectoral business cycles, using the ``Main'' sector-specific elasticities $\hat \theta_i$, and the ``Uniform'' elasticity $\bar \theta$. The text box reports the average response.}}
\end{figure}

\end{document}